\documentclass{article}%
\usepackage{amsfonts}
\usepackage{amsmath}
\usepackage{amssymb}
\usepackage{graphicx}%
\begin{document}

\title{Borel Quantisation\\and\\Nonlinear Quantum Mechanics.
\\A Review of Developments in the Series \textquotedblleft Symmetries in
Science\textquotedblright\ I -- XIII}
\author{H.-D. Doebner\\Department of Physics and Material Sciences
\\D--538670 Clausthal-Zellerfeld\\email: asi@pt.tu-clausthal.de
\and J. Tolar\\Faculty of Nuclear Sciences and Physical Engineering
\\Czech Technical University\\CZ--115 19 Prague, B\v{r}ehov\'a 7
\\email: jiri.tolar@fjfi.cvut.cz}
\date{}
\maketitle

\section{Introduction}

In the first Symmetries in Science meeting in 1979 at Carbondale we presented
preliminary results for a quantisation of the classical kinematic for
non--relativistic systems which are localised and moving on a smooth manifold
$M$. Our paper \cite{1} `On Global Properties of Quantum Systems'
was published in the Proceedings of Symmetries in Science series.
We developed subsequently (with Bernd Angermann) \cite{2,3}
a quantisation method on smooth manifolds --- the `Quantum Borel Kinematics'
(QBK); for a recent review see \cite{4}. In 1992 a suitable time dependence
was proposed (with Jerry Goldin) (see the review \cite{5}
and a more general `Borel Quantisation' (BQ) which emerged from geometrical
and topological considerations; it indicated a nonlinear extension of quantum
mechanics. We participated in some of the later editions of Symmetries in
Science series, often together with members of the `Clausthal group', e.g.
Vlado Dobrev, Jerry Goldin, Wieland Groth, J\"{o}rg Hennig, Wolfgang
L\"{u}cke, Hans-J\"{u}rgen Mann, Peter Nattermann, Wolfgang Scherer,
Christoph Schulte, Pavel \v{S}\v{t}ov\'{\i}\v{c}ek and Reidun Twarock.
The results, different aspects and applications of
Borel Quantisation can be found in the volumes of `Symmetries in Science'.

Our interest in quantum mechanics on manifolds was connected with the
following situation: During 1970--1980 some of our colleagues in quantum theory
and in particle physics thought that Lie groups and their representations
are a major key to model and to understand particle physics.
In this context we worked e.g. on spectrum generating algebras and on
embeddings of physical Lie algebras. Based on Mackey's theory of induced
representations we wrote a paper \cite{6} on a quantisation of particles
moving on homogeneous $G$--spaces. We realised that the geometry
of the $G$--space does not contain `enough' information for a time evolution
on $G$. Furthermore, we failed to generalise Mackey's method to physical
important non--homogeneous spaces. Group theory was obviously a very successful
model, but it was too `rigid': If one chooses the group
and its representation, the complete mathematical framework is already given;
there does not appear the flexibility which one wants for a description
of physical systems. Hence those mathematical formalisms which are `close' to
group theory and which are in addition more `flexible' became interesting.
Among such formalisms are: nonlinear and non--integrable representations
of Lie algebras and their deformations in the sense of Gerstenhaber.
A further promising field for a geometric modelling are differential
geometrical and algebraic notions on $M$. Here one views physical laws
e.g. as relation between geometrical or algebraic objects living on $M$.
Following the pioneering papers of George Mackey \cite{7} and
Irving Segal \cite{8}, we found a path to understand quantisations
of a system on a topologically nontrivial configuration space and how the
quantised system `feels' the topology. This leads to Quantum Borel
Kinematics characterised by topological quantum numbers and one
additional quantum number $D$. This $D$ is connected with the structure
of the infinite--dimensional Lie algebra spanned by quantised kinematical
operators.

In our kinematical design a physical interpretation of $D$ was obscure.
It became more transparent in 1988 when Jerry Goldin and one of the authors
(HDD) realized that two approaches are equivalent: the quantised Borel
kinematics on Euclidean configuration spaces and the representations of
non--relativistic current algebras on multiparticle configuration spaces for
indistinguishable objects found by Goldin and co-workers \cite{9}.
The quantum number $D$ appears in front of an additional term
in the generalised momentum operator as well as in the momentum current.
Jerry Goldin and HDD introduced, based on this observation,
a generic time dependence for pure states and derived
a family of nonlinear Schr\"{o}dinger equations \cite{10} --- DG equations
--- with nonlinear term proportional to $D$. Special generalisations
to mixed states (von Neumann equations) are known \cite{11}.
A direct connection of the DG family to certain nonlinear gauge
transformations \cite{12,13} and an interpretation of $D$ through
nonlinear transformation \cite{14} was elaborated.

Some of these developments are reviewed and commented in this contribution.

\newpage

\section{Borel Kinematic}

\subsection{Classical Case}

We start with a classical model for the kinematic of a system (particle)
localised and moving on a smooth manifold $M$. The kinematic is characterised
by the following set of observables:

\begin{quote}
{\it Generalised positions}
\[
f\ \in C^{\infty}(M,R)
\]
realised by real functions on $M$, and 
\\ {\it generalised momenta}
\[
X\in Vect(M)
\]
realised by smooth vectorfields on $M$.
\end{quote}
We define the tuple
\[
S(M)=(C^{\infty}(M,R),Vect(M))
\]
as the generic KINEMATIC on $M$ (or covariance algebra of $M$).
$S(M)$ has the following properties:
\begin{enumerate}
\item $Vect(M)$ is a ($\infty$--dimensional) Lie algebra of a subgroup
of the diffeomorphism group DIFF$(M)$ of $M$;
\item $C^{\infty}(M,R)$ can be viewed as 
an ($\infty$--dimensional) Abelian
Lie algebra;
\item $f$ and $X$ defined on $M$ form a semidirect sum
\[
S(M)=C^{\infty}(M,R) \oplus_{s} Vect(M).
\]
\end{enumerate}
For physical reasons (see section 2.3) we restrict $Vect(M)$
to the  subset of complete vectorfields $Vect_{0}(M)$; 
this subset spans a partial Lie algebra) 
(the Lie bracket of two complete vector fields may not be
complete) and the corresponding kinematic 
$S_{0}\left(  M\right)$ has partial Lie algebra structure.

\newpage

\subsection{Quantisation of $S_{0}(M)$}

To quantise the classical object $S_{0}(M)$ we construct a map from
$S_{0}(M)$ into the set of essentially selfadjoint operators on a common dense
domain in a separable Hilbert space $H$%
\[
\mathbb{Q}=(\text{\textsf{Q}},\mathsf{P}): \quad 
S_{0}(M) \longrightarrow  SA(H)
\]
sending
\[
f\longrightarrow \textsf{Q}(f), \quad
X\longrightarrow \mathsf{P}(X)
\]

We realise $H$ as $L^{2}(M,\mathbb{C},d\nu)$, i.e. the space of
 square integrable complex functions over $M$; $d\nu$ is a standard measure
on $M$. We assume furthermore that there is no internal degree of freedom
like spin. The map can be viewed as a
representation of an infinite dimensional (partial) Lie algebra.

The following properties I.-III. are assumed for $\mathbb{Q}$
to be a {\it quantisation map}:
\begin{quote}
{\bf I.} In $L^{2}(M,\mathbb{C},d\nu)$ operators $\textsf{Q}(f)$ act as
multiplication operators by $f$,
i.e.
\[
\textsf{Q}(f)\Psi=f\Psi.
\]

{\bf II.} The Lie algebra structure of $S_{0}(M)$ survives,
i.e. $\mathbb{Q}$ is a partial Lie algebra homomorphism.

{\bf III.} $\mathsf{P}(X)$ is a local operator, i.e.
$supp \ (\mathsf{P}(X)\Psi)\subset supp \ \Psi$.
\end{quote}

These assumptions have the following background:

Ad {\bf I.}
Consider a set of localisation regions $B\subset M$; choose for this set the
Borel field $\mathfrak{B}(M)$ over $M$ and define the quantisation map:
\[
\mathbb{Q} : B \in  \mathfrak{B}(M) \longrightarrow E(B)\in SA(H).
\]
The states of the system are given by normed positive 
trace--class operators $W$.
The expectation value of a measurement of $E(B)$ in a state $W$ is
\begin{equation} \label{1}
Tr(WE(B))=\mu_{W}(B)
\end{equation}
($Tr(.)$ denotes trace) and contains information on the probability
of localisation of the system in state $W$ in the region $B$.
Using properties of position measurements we assume that the
r.h.s. of (\ref{1}) is (elementary) spectral measure on $\mathfrak{B}(M)$.
With our realisation of the spectral theorem we find property {\bf I.}

Ad {\bf II.}
The algebraic structure of $S_{0}(M)$ reflects that the classical system is
localised and moving on $M$. Also the quantum system lives on $M$. Hence this
algebraic structure should `survive' under the quantisation map. In this sense
quantisations are based ``on an algebra'' \cite{15}.

Our later analysis shows that there are different possibilities for such maps
which are related to unitarily inequivalent quantisations. Therefore the
information encoded in the classical system is not sufficient to characterise
its quantised form; one needs additional information ---
so called `quantum information'.

Ad {\bf III.}
For $M=R^{n}$ we know that the momentum operator acts in $H$ as a selfadjoint
differential operator. It is plausible to expect also that \textsf{P}$(X)$
acts locally on complex functions over $M$ as differential operator.
Hence in our design we have to define differential operators (of finite order)
on functions in $L^{2}(M,\mathbb{C},d\nu)$. For this we sketch two notions
({\bf A, B}) and their relation ({\bf C}):

{\bf A}.
To define derivatives of complex functions over $M$ one needs a differentiable
structure $DS$ on the point set $M\times\mathbb{C}$. The restrictions of $DS$
give the differentiable structure $DS(M)$ of $M$ (smooth manifold)
\[
 DS(M\times\mathbb{C})|_{M}=DS(M)
\]
and the restriction to $\mathbb{C}$ yields the standard
 differentiable structure of $\mathbb{C}$,
\[
DS(M\times\mathbb{C})|_{\mathbb{C}}=DS(\mathbb{C})
\]

Geometrical objects with these properties are complex line bundles on $M$
with hermitian connection,  $\mathfrak{L}(M\times\mathbb{C},pr,\mathbb{C},
\left\langle . \right\rangle ,\nabla)$. Some sections of $\mathfrak{L}$ are
square integrable. The Hilbert space $L^{2}(M,\mathbb{C},d\nu)$ can be viewed
as the space of square--integrable sections of $\mathfrak{L}$.

The structure of the set $\left\{  \mathfrak{L} \right\}$ 
of such bundles is known. Denoting the curvature of the connection $\nabla$ in $\mathfrak{L}$ by $R$,
one can construct such $\mathfrak{L}$ if and only if
\[
\frac{1}{2\pi i} \int_{S} R  \; \in \; \mathbb{Z}
\]
for all closed 2-surfaces $S$ in $M.$  In terms of cohomology the
de Rham class of $\frac{1}{2\pi i}R$ has to be integral,
\begin{equation}
\left[  \frac{1}{2\pi i}R\right]  \in H^{2}(M,\mathbb{Z}),
\label{d2}
\end{equation}
i.e. there is a strong bundle isomorphism between two complex line bundles,
if and only if their Chern classes in $H^{2}(M,\mathbb{Z})$ coincide.

For each of these inequivalent classes there is a set of
inequivalent connections labelled by
\[
H^{1}(M,U(1))=\pi_{1}^{\ast}(M),
\]
i.e. by elements of the character group of the fundamental group of $M$.

These algebraic invariants classify the line bundles together
with their differentiable structures and covariant derivatives $\nabla$,
we are looking for. We have no result whether
the introduction of differentiable structures to our model via line
bundles is unique. For internal degrees of freedom
complex vector bundles can be used \cite{16,17}.

{\bf B.}
For physical reasons we want to avoid nonlocal effects, i.e. we quantise the
kinematic by local operators \textsf{Q}$(f)$ and \textsf{P}$(X)$.
The position operators are local by construction;
the locality of \textsf{P}$(X)$ is equivalent to condition {\bf III.}
Note that we assume locality only for $\mathbb{Q}S_{0}(M)$;
other operators representing observables could be nonlocal.

{\bf C.}
The locality condition is linked with differential operators defined via
differentiable structures \cite{18}: if there is a differentiable structure
$DS(M\times\mathbb{C})$, then the locality of \textsf{P}$(X)$ implies that
\textsf{P}$(X)$ is a differential operator of finite order with respect to
$DS(M\times\mathbb{C})$.

The arguments in section 2.2 are related to our generalisation of Mackey's
imprimitivity theorem on homogeneous spaces. Our review \cite{4}
and references therein have utilised such a generalisation.

\subsection{A Classification Theorem for Quantisation Maps}

With the assumptions {\bf I. -- III.} we derived a classification theorem
for the quantisation maps

\subsection*{Theorem}
{\it 
The set $\{\mathbb{Q}\}$ of unitarily inequivalent quantisation maps
\[
\mathbb{Q}^{(.)}:\ \ \ S_{0}(M)\longrightarrow SA(H)
\]
on a Hilbert space $H$ which is realised with square--integrable sections
of a hermitian line bundle, $\mathbb{Q(}S_{0}(M))$,
 is labelled by the triple
\[
(J,\alpha,D) \; \in \; H^{2}(M,\mathbb{Z})\times\pi_{1}^{\ast}(M)\times R.
\]
Here $H^{2}(M,\mathbb{Z})$ labels the set of closed 
two--forms $J$
which satisfy the integrality condition (\ref{d2}),
$\pi_{1}^{\ast}(M)$ denotes the character group of
the fundamental group of $M$. For a fixed $J$ we have classification by
($\alpha,D).$ }
\vskip 5mm

Explicit form of the map $\mathbb{Q}^{(\alpha,D)}=$ $($\textsf{Q}
$^{(\alpha,D)}),$\textsf{P}$^{(\alpha,D)})$ can be found in \cite{3,4,21}.
Details for the case $M=R^{3}$ are given in section 2.4.

The operators \textsf{Q}$^{(J,\alpha,D)}(f)$ and \textsf{P}$^{(J,\alpha,D)}(X)$ are
essentially selfadjoint (we used complete vector fields)
 on a common invariant domain; the representation $\mathbb{Q}^{(.)}(S_{0}(M))$ is irreducible;
$\alpha$ is a topological quantum number; $D$ is independent of the topology
and is related to the algebraic structure of $S_{0}(M)$. Hence there are
inequivalent quantisations for systems on topologically
 trivial manifolds.

\subsubsection*{REMARKS}
\begin{enumerate}
\item
Quantum Borel kinematics are based on a classical configuration space in
contrast to geometric (pre--)quantisation \cite{19} which works on a
symplectic space or more specially on the phase space. In both methods the
topological quantum numbers play (with different motivations) an
essential role. However, the quantum number $D$ appears only in quantum Borel
kinematics. This gap was closed recently: J\"{o}rg Hennig and Peter
Nattermann showed in \cite{20,21} that geometric quantisation of the kinematic
corresponds to our approach, if one uses $(\frac{1}{2}-i\gamma)$--density
instead of a $\frac{1}{2}$--density; the imaginary part of the density is
proportional to the quantum number $D$.
\item
We sketch in paragraphs A, B, C formulations of quantum Borel kinematic
for more general situations:
\begin{itemize}
\item[A.]
If the system has internal degrees of freedom like spin, the Hilbert space is
spanned by vector--valued functions. There are two types of quantum maps:
type $0$ in which different vector components are not mixed; this type is
described in \cite{16}. In type 1 a mixing is allowed; Michael Drees gave some
preliminary results \cite{17}. A discussion of the quantum Borel kinematic
for a spinning particle can be found in \cite{22,23}.
\item[B.]
If there exists an external field or potential on $M$,
i.e. a closed 2-form $B$, an additional term in the commutator
of the momentum observables appears.
Hence a quantisation map based on $M$ exists only if $B$ fulfils
the integrality (admissibility) condition (\ref{d2}).
We refer to \cite{4,21}. However, such
external potentials were already included in an indirect way in our earlier
discussion, since the quantum number $J$ is responsible for the existence
of both the line bundles and the admissible closed external one--forms.
\item[C.]
The physical background of $D$ cannot be completely clarified in a kinematical
framework. One can calculate how expectation values of momenta depend on
$D$. But this dependence can be analysed only if it is explicitly known.
We discuss this in section 3.
\end{itemize}
\end{enumerate}

\subsection{Applications of the Classification Theorem}

The theorem shows that for  topologically trivial as well as non--trivial
manifolds `different quantisations' exist. This means that the probabilities
of certain observables measured in certain states depend on $(J,\alpha,D)$.
One and the same classical system yields --- after 
the quantisation maps --- a set of different quantum
systems. Additional information --- the already mentioned
`quantum information' --- is necessary to choose
or to determine $(J,\alpha,D).$ The source of this
information can be, e.g., first principles or experimental results.

\begin{table}[t]
\caption{Examples of elementary quantum Borel kinematics
\cite{4}. \label{tab}}
\vspace{0.2cm}
\begin{center}
\footnotesize
\begin{tabular}{|l|c|c|c|c|c|}
\hline
{Quantum system} & \raisebox{0pt}[13pt][7pt]{$M$} &
\raisebox{0pt}[13pt][7pt]{$\pi_{1}(M)$} &
{$H_{1}(M,Z)$} & {$H^{2}(M,Z)$}&
\begin{minipage}{1in}
\begin{center}
Topological \\ quantum \\ numbers
\end{center}
\end{minipage}
\\[5pt]
\hline
Spinless particle in $R^{3}$   & $R^{3}$ & $\{e\}$ &
$0$  & $0$ & ---  \\[5pt]
Aharonov--Bohm configuration & $R^{3} \backslash R $ &
$Z$ & $Z$ & $0$ & $\vartheta \in [0,1)$ \\[5pt]
Dirac's monopole & $R^{3} \backslash O = R_{+} \times
S^{2}$ & $\{e\}$ & $0$ & $Z$ & $n \in Z $  \\[5pt]
\begin{minipage}{1.5in}
2 distinguishable \\ particles in $R^{3}$
\end{minipage}
& $R^{3} \times R_{+} \times S^{2}$ & $\{e\}$  & $0$
& $Z$ & $n \in Z $
\\[10pt]
\begin{minipage}{1.5in}
2 indistinguishable \\ particles in $R^{3}$
\end{minipage}
& $R^{3} \times R_{+} \times RP^{2}$ & $S_{2}$  & $Z_{2}$
& $Z_{2}$ & $m \in Z_{2} $
\\[10pt]
Rigid body & $R^{3} \times SO(3)$ & $Z_{2}$  & $Z_{2}$
&$Z_{2}$ & $m \in Z_{2} $
\\[5pt]
Symmetric top & $S_{2}$ & $\{e\}$ & $0$ & $Z$ & $n \in Z $
\\[5pt]
Rotator with fixed axis & $S^{1}$ & $Z$ & $Z$ & $0$
& $\vartheta \in [0,1) $
\\[5pt]
\begin{minipage}{1.5in}
Particle on orientable \\ surface of genus $p$
\end{minipage}
& $K_{p}$ & $\pi_{1}(K_{p})$ & $Z^{2p}$ & $Z$ &
\begin{minipage}{1in}
\begin{center}
$n \in Z, $ \\ $\vartheta_{1} \ldots \vartheta_{2p}
\in [0,1)$
\end{center}
\end{minipage}
\\[10pt]
\hline
\end{tabular}
\end{center}
\end{table}

As the first example of inequivalent quantisations we consider a topologically
trivial manifold $M=R^{3}$ with vector fields
$X(\vec{g})=\vec{g}(\vec{x}) . \vec{\nabla}$
and with quantisation map $\mathbb{Q}^{(D)}(S_{0}(R^{3}))$,
\begin{align}
\mathsf{Q}(f)  &  =f    \nonumber\\
\mathsf{P}^{(D)}(X)  &  =-i\hbar\vec{g} . \vec{\nabla}+
(-i\frac{\hbar}{2}+D)\; div \ \vec{g} \label{3}
\end{align}
Different $D$ yield unitarily inequivalent representations and hence different
quantum systems.

The representations of $S_{0}(R^{n})$ for multiparticle configuration spaces
for $N$ indistinguishable objects can be viewed also as representations of
non--relativistic inhomogeneous current algebras.
Jerry Goldin and co-workers \cite{9} constructed such representations;
they derived the above result (\ref{3}) and found
independently the quantum number $D$.

Now let us mention some systems on topologically non--trivial
smooth manifolds and their different quantisations
(see also Table 1):

The physics of $N$ indistinguishable particles --- 
anyons ---
and distinguishable particles on a 2--dimensional Euclidean space
(in the framework of current algebra)
was discussed by Jerry Goldin and co-workers. A review (in Borel quantisation)
of indistinguishable and distinguishable particles on 2--dimensional
manifolds can be found in \cite{24}. Parastatistics appears for $N$
indistinguishable particles on manifolds with dimension $>3$ \cite{24}.
Aharonov--Bohm situations were discussed as topological effects in
\cite{25}. The quantum maps for systems on non--orientable
2--dimensional manifolds
(M\"{o}bius strip and Klein bottle) were treated in \cite{26}.
For quantisations on the trefoil knot see \cite{27}.
A recent review of many aspects in quantum Borel kinematics can be found in
\cite{4} together with some further examples of configuration spaces with
nontrivial topology.

\newpage

\section{Borel Dynamics}

\subsection{Difficulties with $\mathbb{Q}^{(.)}(S_{0}(M))$}

Quantum Borel Kinematic holds for any fixed time $t$, it considers a `frozen'
system and it carries no direct information on a 
$t$--dependence. Hence a
principle is needed to construct related dynamical equations. As explained
before one hopes that this could be a key for a physical interpretation of the
quantum number $D$.

As a plausible model for an evolution of pure states 
we choose a dynamical group
$\{\mathfrak{D}_{t,g};t\in R\}$ with a linear operator $\mathbb{D}_{g}$ as
generator acting on $H$; $g$ denotes a Riemannian structure on $M$. 
In the Heisenberg picture we relate the quantised momentum \textsf{P}$^{(.)}(X)$ to
$\mathbb{D}_{g}$, i.e. we assume the existence of a map
\[
C^{\infty}(M,R)\rightarrow Vect_{0}(M)
\]
such that
\[
\left[  \mathbb{D}_{g},\text{\textsf{Q}}(f)\right]  =-i\mathsf{P}^{(.)}
(X_{f})  \quad  \text{for all} \quad f\in C^{\infty}(M,R).
\]
In analogy to Hamiltonian mechanics in phase space we specialise this map with
\[
X_{f}= grad_{g} \ f .
\]
To analyse this ansatz consider the  $\nabla$--lift of the Laplace Beltrami
operator $\Delta_{g}$ on $M$ (with metric $g$) is a candidate for $\mathbb{D}$.
Hence we write (with some operator 
$\mathbb{K}=\mathsf{Q}(V)$)
\[
\mathbb{D}_{g}=-\frac{1}{2}\Delta_{g}^{\nabla}+\mathbb{K}.
\] 
The commutator between $\mathbb{D}_{g}$ and $f$ has 
with $(\ref{3})$ for all $f \in C^{\infty}(M,R)$ 
the form 
\[
\lbrack\mathbb{D}_{g},f]=i \mathsf{P}(grad_{g} \ f) -
       i D \Delta_{g} f.
\]
A comparison with the previous result yields $D=0$.
Thus our ansatz for a dynamical group
 (in the Heisenberg picture) fails; it leads to the trivial result $D=0$ \cite{28}. This failure is partly connected
 with the implicit assumption that evolutions of wave
 functions are linear.

\subsection{Nonlinear evolutions from $\mathbb{Q}^{(D)}(S_{0}(R^{3}))$}
An alternative method is to start with the assumption that the positional
probability is conserved. Consider again a system in $M=R^{3}$. We assumed [5]
\[
\frac{\partial}{\partial t}\int_{R^3} \varrho(x,t)d^{3}x=0,\text{ \ \ \ \ \ \ }%
\varrho(x,t)=\overline{\Psi}(x,t)\Psi(x,t).
\]
This implies indirectly that a pure state remains a pure state. With a
suitable behaviour of $\Psi$ at infinity we are allowed to apply the Gauss
theorem and find
\begin{equation}
\frac{\partial}{\partial t}\varrho(x,t)=-\nabla\vec{j}(x,t) \label{b}
\end{equation}
with a vector field density $\vec{j}$ depending on the wave function $\Psi$.

How to construct $\vec{j}$ in our model? Consider the above equation as an
operator equation in the 1--particle sector $\mathfrak{F}_{1}$ of the Fock
space generated from a cyclic vacuum $\left\vert 0\right\rangle .$ The generic
operators  \textsf{Q}$(f)$, \textsf{P}$^{(D)}(X)$ correspond to operator--valued
densities $\varrho$, $\vec{j}^{D}$ in $\mathfrak{F}_{1}$:
\begin{align*}
  \mathsf{Q}(f) &= \int f(x)\varrho(x,t)d^{3}x\\
\mathsf{P}^{(D)}(X)  &  = \int\vec{g}(x)\vec{j}^{D}(x,t)d^{3}x
\end{align*}
We have already $\varrho(x,t)=\overline{\Psi}\Psi.$ 
We get for $\vec{j}^{D}(x,t)$ from (\ref{3})
\[
\vec{j}^{D}(x,t)=\vec{j}^{0}(x,t)-D\nabla\varrho(x,t),
\text{ \ \ }
\vec{j}^{0}(x,t)=\frac{\hbar}{2mi}(\overline{\Psi}
\cdot \nabla \Psi - \overline{\nabla\Psi}\cdot\Psi)
\]
and with $(\ref{b})$
\[
\frac{\partial}{\partial t}\varrho(x,t)=-\nabla\vec{j}^{0}(x,t)+D\nabla
\varrho(x,t).
\]
This is a Fokker--Planck type equation. For $D=0$ we have the quantum
mechanical continuity equation with the usual quantum mechanical current. The
term proportional to the quantum number $D$ is a quantum mechanical diffusion
current which is characteristic for the model.

Any ansatz for a time dependence of $\Psi(x,t)$ has to respect this Fokker
Planck equation. We use this fact to construct an evolution of first order in
$\partial_{t}$, with the usual linear terms and with an additional term $F$
depending e.g. on the wave function $\Psi$:
\[
i\hbar\frac{\partial}{\partial t}\Psi(x,t)=(-\frac{\hbar^{2}}{2m}%
\Delta+V(x)+F\left[  \Psi\right]  \Psi
\]
Inserting this ansatz into the Fokker--Planck equation,
a non linear Schr\"{o}dinger equation with a complex nonlinear term is obtained,
\[
i\hbar\frac{\partial}{\partial t}\Psi(x,t) =
(-\frac{\hbar^{2}}{2m}
\Delta+V(x)+iIm \ F \left[  \Psi\right]  +Re \ F \left[  \Psi\right]  )\Psi,
\]
where
\[
Im \ F \left[  \Psi\right]    =
\hbar\frac{D}{2}\frac{\Delta\varrho}{\varrho}
\]
 is enforced through $\mathbb{Q}^{(D)}$, and
$Re \ F\left[  \Psi\right]$ is independent of $D$ (arbitrary).

We see that the imaginary part of $F$ is fixed by the quantisation method, but there is
no information on the real part. We assume for Re F a function of the wave
function which is of the same type as Im$F$, i.e.

\textbullet\ complex homogeneous of order zero

\textbullet\ rational with derivatives not higher than second order in the numerator

\textbullet\ Euclidean invariant.

With these assumptions for $Re \ F$ Doebner and Goldin obtained a family of
singular nonlinear Schr\"{o}dinger equations (DG equations) for a particle
with mass $m,$ potential $V$ parametrised by
$\hbar$, $D$, $c_{1}$,..., $c_{5}$:
\begin{align*}
i\hbar\frac{d}{dt}\Psi &  =\left[  -\frac{\hbar^{2}}{2m}\Delta+V(\vec
{x})+\frac{1}{2}\hbar D\frac{\Delta\varrho}{\varrho}+\hbar D^{\prime}%
\sum_{i=1}^{5}c_{i}R_{i}\left[  \Psi\right]  \right]  \Psi\\
R_{1}\left[  \Psi\right]   &  =\frac{m}{\hbar}\frac{\nabla\vec{j}^{0}}%
{\varrho},\text{ \ \ }R_{2}\left[  \Psi\right]  =\frac{\Delta\varrho}{\varrho
},\text{ \ \ }R_{3}\left[  \Psi\right]  =\frac{m^{2}}{\hbar^{2}}\frac
{(\nabla\vec{j}^{0})^{2}}{\varrho^{2}}\\
R_{4}\left[  \Psi\right]   &  =\frac{m^{2}}{\hbar^{2}}\frac{(\nabla\vec{j}%
^{0})^{2}}{\varrho^{2}},\text{ \ \ }R_{5}\left[  \Psi\right]  =\frac
{(\nabla\varrho^{2})}{\varrho^{2}}.
\end{align*}
The choice of the real nonlinearity corresponds to the `gauge generalisation'
used in another derivation of the DG equations (see sections 4.1, 4.2).

Independently of the known fact (see e.g. the review 
\cite{5})
 that the usual framework of quantum mechanics does not allow fundamental
nonlinear evolutions for pure states, one may argue
 that a small nonlinearity
(small $D$) can be treated approximately with the usual methods. With this
precaution some results for atomic spectra were derived; for the hydrogen
atom present precession experiments show no difference to the linear
theory. This leads to an upper bound for $D$ \cite{29}
\[
D>10^{-7}\frac{\hbar}{m}%
\]
There are discussions on DG type nonlinearities in quantum optics (`nonlinear photons') \cite{29}.

For an exact calculation of observable effects a new 
framework of quantum mechanics is necessary. There are
indications how to formulate general
requirements (e.g. \cite{30}), but there is by no means a complete and mathematically
acceptable theory which incorporates fundamental nonlinearities.

The mathematical structures and properties of DG equations and the mentioned
precaution for their physical applications are partly known; we quote (the
following list is incomplete):

Cauchy problem \cite{31}; Lie symmetries \cite{32};
solutions for stationary states \cite{10}, time dependent
solutions for certain coefficients \cite{33};
generalisations for: arbitrary smooth $M$ \cite{21}, 
mixed states \cite{11}; other methods for a
derivation via: nonlinear gauge transformations \cite{13},
generalised Ehrenfest relations \cite{11}, 
stochastic processes \cite{34}. Further
applications are known for anti--particles \cite{35} and 
the dynamics for $D$--branes.

\newpage

\section{Borel Kinematic and Nonlinear Structures}
Our quantum number $D$ yields a family of nonlinear evolution equations for pure
states with a nonlinearity proportional to D. This indicates a hidden
nonlinear structure in $\mathbb{Q}^{(0)}(S_{0}(R^{3}))$. We assume in the
following $M = R^3$.

\subsection{Nonlinear Gauge Transformations}

In connection with the properties of stationary solutions of the DG family a
group $\mathcal{G}$ of nonlinear gauge transformations was introduced \cite{12}.
They are invertible transformations $\mathbb{N}$%
\begin{equation}
\mathbb{N\,}: \; \Psi \in H \longrightarrow 
\mathbb{N}\Psi =\mathbb{N}\left[  \Psi\right]  \in H \label{c}
\end{equation}
with $\mathbb{N}$ depending on $\Psi(x,t)$, $x$, $t$. 
They are restricted by the assumption that the positional
probability density is invariant:
\[
\overline{\mathbb{N}\left[  \Psi\right]  }\mathbb{N}\left[  \Psi\right]
=\overline{\Psi}\Psi.
\]
The reason for this restriction is the following:
$\mathbb{N}$ should transform a
given system, i.e. a given $\Psi$, to a 
`physically equivalent' one: ``equivalent'' means that 
the results of measurements on both systems are the same.

Behind the notion of `physical equivalence' is the `principle' that the
positional density $\varrho(x,t)$ for all $x$ and $t$ determines the outcomes
of the measurements of all observables \cite{36}.
Such $\mathbb{N}$ build a nonlinear gauge group $\mathcal{G}$.

Now applying $\mathbb{N}$ to the usual (linear) Schr\"{o}dinger equation, i.e. to
a system with linear evolution operator
\[
\mathbb{D}_{S}=i\hbar\partial_{t}+\frac{\hbar^{2}}{2m}\Delta-V(x),\text{
\ \ }\mathbb{D}_{S}\Psi=0,
\]
a (family of) nonlinear Schr\"{o}dinger equations is obtained,
\[
\mathbb{D}_{S}\circ\mathbb{N}\left[  \Psi\right]  =0.
\]
The result is a subfamily in the DG family.

By construction it describes the physics of a system which is equivalent to
the linear system. A generic procedure (gauge generalisation) to construct
`new' systems is the gauge generalisation. This is a generic procedure for
families of partial differential equations depending on coefficients which
are related to each other; the breaking of this relation is our model for 
a ``gauge generalisation''
\cite{12,13}. This gauge generalisation leads to the DG family; some of its
members are inequivalent, they represent systems with new physical properties.

Sections 3.2 and 4.1 show that the DG equations can be
derived from two different structures:

\textbullet\qquad From a geometric structure via a representation of an
inhomogeneous diffeomorphism group acting on the configuration space and with
a $t$--dependence from a conservation of positional probability density.

\textbullet\qquad From a nonlinear structure via nonlinear transformations
of positional probability densities between physically equivalent systems
applied to a $t$--dependence of the linear system and gauge generalisation.

\newpage

\subsection{Nonlinear Tangent Map}
Another method to describe a hidden nonlinear structure of $\mathbb{Q}^{(0)}(S_{0}(R^{3}))$ more directly was recently presented \cite{14}.

Let $\mathbb{N\in}\mathcal{N}$ as in (\ref{c}) be a group of nonlinear
transformations in $H$. To introduce convenient transformation properties
for operators we consider physically interesting ones which are (often)
essentially selfadjoint. They can be viewed as generators
$i\mathbb{A}$ of a one parameter group $U_{\varepsilon}$ of
unitary transformations
\[
U_{t}=\exp it\mathbb{A}.
\]
Take a path $\{U_{t} \Psi, \ t \in
\lbrack - \varepsilon,\varepsilon]\}$ in $H$.
Then
\[
\frac{d}{dt}(U_{t}\Psi)|_{t=0}=
\frac{d}{dt}(\Psi+it\mathbb{A}\Psi)|_{t=0}=
i\mathbb{A}\Psi.
\]
Hence $i\mathbb{A}\Psi$ appears as a tangent map $T$ of
$U_{\varepsilon}$. Take now a transformed path
$\mathbb{N}(U_{\varepsilon})$ and define the transformed
generator $i\mathbb{A}_{\mathbb{N}}$ by the tangent map
$T(\mathbb{N})$ of $\mathbb{N}(U_{\varepsilon})$:
\[
\frac{d}{dt}\mathbb{N}(U_{\varepsilon}\Psi)|_{t=0}=
i\mathbb{A}_{\mathbb{N}}\Psi.
\]
Hence we have the $\mathbb{N}$-tangent map
\[
T(\mathbb{N}):\text{ \ \ }\mathbb{A\rightarrow A}_{\mathbb{N}}.
\]
For linear $\mathbb{N}$ we get the usual result. For nonlinear $\mathbb{N}$
the resulting operator $\mathbb{A}_{\mathbb{N}}$ is in general nonlinear. The
$\mathbb{N}$-tangent map is a Lie algebra isomorphism. One can extend the
method formally to non essentially selfadjoint operators.

We apply now the $\mathbb{N}$-tangent map to quantise
kinematical observables. Similarly as in section 4.1 
we restrict $\mathbb{N}$ such that the
$\mathbb{N}$-tangent mapped elements in $\mathbb{Q}^{(D)}(S_{0}(R^{3}))$ are
again linear and of order 0 or 1, i.e.

1.\qquad\textsf{Q}$(f)$ is a linear multiplication operator, i.e. $f$

2. \ \ \ \ \ \textsf{P}$^{(D)}(X)$ is a linear differential operator of order 1.

Condition 1. is fulfilled by construction; condition 2. is
equivalent to the relation
\[
\mathsf{P}_{\mathbb{N}}^{(0)}(\vec{g} \cdot \nabla)=
\vec{g}_{1} \cdot \nabla+g_{0}
\]
with $\vec{g}_{1}(x)$, $g_{0}(x)$ depending on $\vec{g}(x)$. The last condition
implies [14] a set $\mathbb{N}$ of non linear transformations. We write this
(formally) in polar decomposition%
\[
\Psi=R\exp iS,\text{ \ \ \ \ \ \ \ \ }\mathbb{N}[\Psi]=\mathbb{N}%
[R,S]=R_{\mathbb{N}}(R,S)\exp iS_{\mathbb{N}}(R,S)
\]

The form of $\mathbb{N}$ is
\begin{align*}
R_{\mathbb{N}}(R,S)  &  =R^{\kappa+1}r(S)\\
S_{\mathbb{N}}(R,S)  &  =\gamma\ln R+t(S)+1
\end{align*}
with $\kappa,\gamma\in R$, and real functions $r(S)$ and $t(S)$. The functions
$\vec{g}_{1}$, $g_{0}$ are
\begin{align*}
\vec{g}_{1}  &  =+\frac{\hbar}{i}\vec{g}\\
g_{0}  &  =(\frac{\hbar}{2i}+\frac{1}{4}\gamma) div \ \vec{g}
\end{align*}

The transformations $\mathbb{N}(\kappa,\gamma,r(S),t(S))$ build a group  $\mathfrak{N}$.

\subsection{Applications of the Nonlinear Tangent Map}

We consider the behaviour of $\mathbb{Q}^{(D)}(S_{0}(R^{3}))$ under
$\mathfrak{N}$. For $D=0$, i.e. for $\mathbb{Q}^{(0)}(S_{0}(R^{3}))$, we have (see (\ref{3}))
\begin{align*}
\mathsf{Q}(f)  &  =f\\
\mathsf{P}^{(0)}(X)  &  =\frac{\hbar}{i}
\vec{g} \cdot \nabla-i\frac{\hbar}{2} div \ \vec{g}.
\end{align*}
Hence with $\gamma=4D$
\[
\mathbb{Q}^{(D)}(S_{0}(R^{3}))=\mathbb{Q}_{\mathbb{N}}^{(0)}(S_{0}(R^{3}))
\]
holds. Representations of $\mathbb{Q}^{(D)}(S_{0}(R^{3}))$ with different $D$
which are inequivalent under linear unitary transformations are
`equivalent' with respect to certain non--unitary ones. It is interesting to
apply the $\mathbb{N}$-tangent map to the linear Schr\"{o}dinger equation. For
$\mathbb{N}$ we find that
\[
\mathbb{D}_{S}^{0}\circ\mathbb{N}\left[  \Psi\right]  =0,\text{ \ \ \ \ }
\mathbb{N\in}\text{ }\mathfrak{N}
\]
leads to a subfamily of generalized NLSE which contain (after gauge
generalisation) the DG family. After gauge generalisation a 
``general'' DG family is constructed.

If one applies $\mathbb{N}$ to an ordered polynomial
generated from \textsf{P}$^{(D)}(X),$ \textsf{Q}$(f)$
one gets a nonlinear quantisation
of all observables of polynomial type. The partial 
Lie algebra structure of
these nonlinear operators is known.

\section{Summary and Outlook}

We started with a geometrical framework to quantise the kinematic of a system
living on a topologically nontrivial manifold (Quantum Borel Kinematic). We
showed how the quantisation depends on the topology through topological
quantum numbers. Since inhomogeneous diffeomorphisms are used as models for
the kinematic, a new quantum number of non--topological origin appears.
If a time dependence is introduced to the quantized kinematic through
conservation of probability, this quantum number is the root of
nonlinear Schr\"{o}dinger equations (DG equations) for pure states.
Properties of stationary solutions of
the DG equations lead to the introduction of nonlinear gauge transformations
which transform a system in a physically equivalent one. Applied to the linear
Schr\"{o}dinger equation, nonlinear gauge transformations in $\mathcal{G}$
lead after gauge generalisation to the DG equations.
A more direct indication of an
intrinsic structure of the quantum Borel kinematic utilises
$\mathbb{N}$--tangent maps, $\mathbb{N\in}\mathcal{N}$, which transform
linear quantised kinematical operators into nonlinear ones. These
$\mathbb{N}$--tangent maps can describe a
nonlinear quantisation of polynomial observables; for the Hamiltonian a
generalised family of DG equations appears.

Topological effects in quantum mechanics and the topological quantum numbers
are a well established field with few applications to real systems, e.g.
indistinguishable particles in $R$ , anyons, Bohm--Aharonov situations.
Nonlinear quantum mechanics for pure states is an interesting but
controversial field. On one hand it seems to be plausible that quantum theory
is a linearisation of a more involved theory and that nonlinear evolutions
derived from first principles are a key stone for a formulation
of such a framework. On the other hand we know that linear structures
are deeply rooted in the mathematical and
physical formulation of quantum theory. The present formalism does not allow
nonlinear operators; only for approximations see \cite{37}.
A `new' formalism is not yet developed. There is no experimental
indication for a fundamental nonlinearity. However, deviations from usual
quantum mechanics are discussed in connection with quantum mechanical
precision experiments and with new possibilities for an experimental design.
In this connection topological viewpoints as well as nonlinearities in the
evolution are of interest.

\subsection*{Acknowledgements}
One of the authors (HDD) acknowledges discussions with Alois Kopp.


\begin{thebibliography}{99}

\bibitem{1}
H.-D. Doebner and J. Tolar, in {\it Symmetries in Science,}
eds. B. Gruber and R.S. Millman (Plenum, New York, 1980, 475); in {\it Symmetries in Science II,} eds. B. Gruber and
R. Lenczewski (Plenum, New York 1986, 115)
\bibitem{2}
B. Angermann and H.D. Doebner, Physica 114A, 433 (1982)
\bibitem{3}
 B. Angermann, H.D. Doebner and J. Tolar, 
Lecture Notes in Mathematics, Vol. 137
(Springer, Berlin, 1983, 171)
\bibitem{4}
H.D. Doebner, P. \v{S}\v{t}ov\'{\i}\v{c}ek and J. Tolar,
Rev. Math. Phys. 13, 799 (2001)
\bibitem{5}
H.D. Doebner and G.A. Goldin, `Remarks and Recent Results on
 Nonlinear Extensions of Quantum Mechanics', 
J.Phys. A (in print) (2003)
\bibitem{6}
H.D. Doebner and J. Tolar, J. Math. Phys. 16, 975 (1975)
\bibitem{7}
G.W. Mackey, `Mathematical Foundations of Quantum
Mechanics', 
(W.A.Benjamin, New York, 1963)
\bibitem{8}
I.E. Segal, J.Math. Phys. 1, 468 (1960)
\bibitem{9}
G.A. Goldin, R. Menikoff and D.H. Sharp, 
J. Math. Phys. 21, 650 (1980),
J. Phys. A: Math.Gen. 16, 1827 (1983)
\bibitem{10}
 H.D. Doebner and G.A. Goldin, 
J. Phys. A: Math. Gen. 27, 1771 (1994)
\bibitem{11}
H.D. Doebner and J.D. Hennig, in 
{\it Symmetries in Science VIII,}
ed. B. Gruber (Plenum Press, New York, 1995, 85)
\bibitem{12}
H.D. Doebner and G.A. Goldin, Phys. Rev. A 54, 3764 (1996)
\bibitem{13}
H.D. Doebner, P. Nattermann and G.A. Goldin, 
J. Math. Phys. 40, 49 (1999)
\bibitem{14}
H.D. Doebner and J.D. Hennig, `Quantum Borel Kinematics and
 Nonlinear Quantisations' , in preparation (2004)
\bibitem{15}
H.D. Doebner and J. Tolar, in {\it Symmetries in Science V,} 
eds. B. Gruber, L.C. Biedenharn and H.D. Doebner
(Plenum Press, New York, 1990, 137)
\bibitem{16}
H.D. Doebner and U.A. M\"{u}ller, 
J. Phys. A: Math. Gen, 26, 719 (1993)
\bibitem{17}
M. Drees, `Zur Kinematik lokalisierbarer quantenmechanischer
Systeme unter Ber\"{u}cksichtigung innerer Freiheitsgrade
und \"{a}u\ss erer Felder' ,
Ph.D. Dissertation, TU Clausthal (1992)
\bibitem{18}
J. Peetre, Math. Scand. 7, 211 (1959) and 8, 116 (1960)
\bibitem{19}
 N.M.J. Woodhouse, `Geometric Quantisation',  2nd edition
 (Clarendon Press, Oxford, 1992)
\bibitem{20}
 J.D. Hennig, in `Nonlinear, Deformed and Irreversible Quantum Systems',
eds. H.D. Doebner, V.K. Dobrev and P. Nattermann 
(World Scientific, Singapore, 1995, 155)
\bibitem{21}
P. Nattermann, `Dynamics in Borel Quantisation: Nonlinear
 Schr\"{o}dinger Equations vs. Master Equations',
Ph.D. Dissertation, TU Clausthal (1997)
\bibitem{22}
 H.D. Doebner, R. Zhdanov and A. Kopp, `Nonlinear Dirac
 Equations and Nonlinear Gauge Transformations',
 in `Symmetry in Nonlinear Mathematical Physics',
 Proceedings of the Institute of Mathematics of NAS of
 Ukraine (2004)
\bibitem{23}
 H.D. Doebner and A. Kopp, `Nonlinear Pauli Equations', in preparation (2004)
\bibitem{24}
 H.D. Doebner, W. Groth and J.D. Hennig, 
J. Geom. Phys. 31, 35 (1998); 
H.D. Doebner, P. \v{S}\v{t}ov\'{\i}\v{c}ek and J. Tolar,
Czech. J. Phys. B32, 1240 (1982)
\bibitem{25}
 C. Schulte, in {\it Symmetries in Science X,}
eds. B. Gruber and M. Ramek 
(Plenum Press, New York, 1998, 357)
\bibitem{26}
 C. Schulte, in {\it Symmetries in Science IX,} 
eds. B. Gruber and M. Ramek
(Plenum Press, New York, 1997, 313)
\bibitem{27}
 H.D. Doebner and W. Groth, J. Phys. A 30, L503 (1997)
\bibitem{28}
 B. Angermann, `\"{U}ber Quantisierungen lokalisierter
Systeme -- Physikalisch interpretierbare mathematische Modelle', 
Ph.D. Dissertation, TU Clausthal (1983)
\bibitem{29}
 H.D. Doebner, V.I. Manko and W. Scherer, 
Phys. Lett. A 268, 17 (2000)
\bibitem{30}
B. Mielnik, Commun. Math. Phys 15, 1 (1969)
\bibitem{31}
 H. Teismann, in `Group21', Vol. 1,
eds. H.D. Doebner, P. Nattermann and W. Scherer
 (World Scientific, Singapore 1997, 433)
\bibitem{32}
P. Nattermann, Rep. Math. Phys. 36, 387 (1995)
\bibitem{33}
 A.G. Ushveridze, Phys. Lett. A185, 123 and 128 (1994);
V.V. Dodonov and S.S. Mizrahi, Ann. Phys. 37, 226 (1995);
 Physica A 214, 619 (1995)
\bibitem{34}
 H.J. Mann, Rep. Math. Phys. 44, 143 (1999)
\bibitem{35}
 H.D. Doebner and G.A. Goldin, `Extensions of Quantum Mechanics --- Nonlinear Schr\"{o}dinger Equations for
Particles and Antiparticles', in
Proceedings of the 3rd Symposium `Quantum Theory and Symmetries' (World Scientific, Singapore 2004)
\bibitem{36}
 R.P. Feynman and  A.R. Hibbs, `Quantum Mechanics and Path
 Integral' (McGraw Hill, New York 1965)
\bibitem{37}
 W. L\"{u}cke and P. Nattermann, 
in {\it Symmetries in Science X,} eds. B. Gruber
and M. Ramek (Plenum Press, New York, 1997, 197); \\
N.Gisin, in `Nonlinear, Deformed and Irreversible Quantum
 Systems', eds. H.D. Doebner, V.K. Dobrev and P. Nattermann (World Scientific, Singapore, 1995, 109)
\end{thebibliography}
\end{document}